\documentstyle{article}
\begin{document}
\begin{center}
\Large{\bf  RIEMANN GEOMETRY IN THEORY OF THE \\[2mm]FIRST ORDER
SYSTEMS
OF EQUATIONS  }\vspace{4mm}\normalsize
\end{center}
 \begin{center}
\Large{\bf Valery Dryuma}\vspace{4mm}\normalsize
\end{center}
\begin{center}
{\bf Institute of Mathematics and Informatics AS Moldova, Kishinev}\vspace{4mm}\normalsize
\end{center}
\begin{center}
{\bf  E-mail: valery@dryuma.com;\quad cainar@mail.md}\vspace{4mm}\normalsize
\end{center}
\begin{center}
{\bf  Abstract}\vspace{4mm}\normalsize
\end{center}

     Theory of Riemann extensions of the spaces with constant affine connection
     is proposed to study of the properties of nonlinear the first order systems
    of differential equations.

    As example quadratic system of
differential equations
\begin{equation} \label{dryuma:eq1}
 \frac{dx}{ds}=P(x,y),\quad \frac{d y}{ds}=Q(x,y),
\end{equation}
  where
  \[
P(x,y)=a_0+a_1x+a_2y+a_{11}x^2+a_{12}xy +a_{22}y^2,
\]
\[
Q(x,y)=b_0+b_1x+b_2y+b_{11}x^2+b_{12}xy +b_{22}y^2,
\]
and  $a,~b$ are the parameters, is presented in
homogeneous form and is considered as geodesic of three-dimensional space with constant affine
connection depending on the parameters $a,~b$. After the Riemann extension one get a
 six-dimensional space and its properties in relation to the parameters  are investigated.

   The Lorenz system of  equations
\[
    \frac{dx}{ds}=\sigma(y-x),\quad
  \frac{dy}{ds}=rx-y-xz,\quad \frac{dz}{ds}=xy-bz
\]
after  presentation in homogeneous form is considered as geodesic
equations of four dimensional space with constant affine
connection. Based on the eight -dimensional Riemann extension of a
given type space the
 properties of the Lorenz system are studied.

\section{Introduction}

   The subject of consideration is the first order polynomial  systems
    of differential equations
\begin{equation} \label{dryuma:eq1}
\frac{d x^{i}}{ds}=c^i+a^{i}_{j}x^{j}+b^{i}_{ j k}x^{j}x^{k}
\end{equation}
depending on the parameters $a,b,c$.

    They  play an important role in various branches of modern
mathematics and its applications.

   In particular case of the  system of two equations
$$ \frac{d
x}{ds}=a0+a1x+a2y+a11x^2+a12xy+a22y^2,$$$$\frac{d
y}{ds}=b0+b1x+b2y+b11x^2+b12xy+b22y^2 $$
 there are many unsolved problems.

  The spatial first order system of differential equations
  \begin{equation} \label{dryuma:eq3}
\frac{d x}{ds}=P(x,y,z),\quad \frac{d y}{ds}=Q(x,y,z),\quad
\frac{d z}{ds}=R(x,y,z)
\end{equation}
with the functions  $P,Q,R$ polynomial on variables $x,y,z$
 are still more complicated object for the
studying of their properties.

As example the studying of the Lorenz system of equations
\begin{equation} \label{dryuma:eq4}
\frac{d x}{ds}=\sigma(y-x),\quad \frac{d y}{ds}=rx-y-xz,\quad
\frac{d z}{ds}=xy-bz
\end{equation}
and the R\"ossler system
\begin{equation} \label{dryuma:eq5}
\frac{d x}{ds}=-y-z,\quad \frac{d y}{ds}=x+ay,\quad \frac{d
z}{ds}=bx-cz+xz
\end{equation}
which are the simplest examples of the spatial systems have
chaotic behaviour at some values of parameters represent the
difficult task.
 \section{ From the first order system of equations
to the second order systems of ODE}

   The systems of the first order differential equations are not suitable
   object of consideration from the usually point of Riemann geometry.

   The systems of the second order differential equations in form
\begin{equation} \label{dryuma:eq6}
\frac{d^2 x^i}{ds^2}+\Pi^i_{k j}(x)\frac{d x^k}{ds}\frac{d
x^j}{ds}=0
\end{equation}
are best suited to do that.

   They can be considered as geodesics of the affinely connected space
   $M^k$ in local coordinates $x^k$. The values $\Pi^i_{jk}=\Pi^i_{kj}$ are the coefficients of
   affine connections on $M^k$.

    With the help of such coefficients can
   be constructed curvature tensor and others geometrical objects
   defined on variety $M^k$.

    There are many possibilities to present a given system of the first order of equations
    in the form of (\ref{dryuma:eq6}).

    One of them is a following naive presentation.

    For the system
 \begin{equation} \label{dryuma:eq7}
\frac{d x}{ds}=P(x,y),\quad \frac{d y}{ds}=Q(x,y)
\end{equation}
after differentiation with respect the parameter $(s)$  we get the
second order system of differential equations of the form
(\ref{dryuma:eq6}) $$ \frac{d^2 x}{ds^2}=\frac{1}{P}(P_x \frac{d
x}{ds}+P_y \frac{d y}{ds})\frac{d x}{ds}$$  ,
 $$
 \frac{d^2 y}{ds^2}=\frac{1}{Q}(Q_x
\frac{d x}{ds}+Q_y \frac{d y}{ds})\frac{d y}{ds}. $$

   Such type of the system contains the integral curves of
    the system (\ref{dryuma:eq7}) as part of its solutions and can be considered as
    the equations of geodesics of two dimensional space $M^2(x,y)$ equipment by affine
    connections with coefficients
\begin{equation} \label{dryuma:eq9}
\Pi^1_{11}=-\frac{P_x}{P},\quad \Pi^1_{12}=-\frac{P_y}{2P},\quad
\Pi^2_{12}=-\frac{Q_x}{2Q},\quad \Pi^2_{22}=-\frac{Q_y}{Q}.
\end{equation}

    It is apparent that the properties of the system
    (\ref{dryuma:eq7}) have an influence on geometry of the variety $M^2(x,y)$.

    Remark that a given system  is equivalent the
    second order differential equation
\[
\frac{d^2 y}{dx^2}=\ln(Q/P)_y\left(\frac{d
y}{dx}\right)^2+\ln(Q/P)_x\frac{d y}{dx}
\]
which has the solution in form
\[
\frac{d y}{dx}=\frac{Q}{P}.
\]

    By analogy  the spatial system of the first
    order differential equations can be written.

    As example the Lorenz system of equations is equivalent the
    second order ODE
\[
\frac{d^2
y}{dx^2}-\frac{3}{y}\left(\frac{dy}{dx}\right)^2+\left(\alpha
y-\frac{1}{x}\right)\frac{dy}{dx}+\epsilon x y^4-\gamma y^3-\beta
x^3 y^4-\beta x^2 y^3+\delta\frac{y^2}{x}=0,
\]
where
\[
\alpha =\frac{b+\sigma+1}{\sigma},\quad \beta
=\frac{1}{\sigma^2},\quad \gamma
=\frac{b(\sigma+1)}{\sigma^2},\quad \delta
=\frac{\sigma+1}{\sigma},\quad \epsilon=\frac{b(r-1)}{\sigma^2},
\]
which can be obtained by the elimination of variable $z$ from the
system
\[
\frac{dy}{dx}=\frac{rx-y-xz}{\sigma(y-x)},\quad
\frac{dz}{dx}=\frac{xy-bz}{\sigma(y-x)}.
\]

   The second order ODE's
\[
\frac{d^2
y}{dx^2}+a_1(x,y)\left(\frac{dy}{dx}\right)^3+3a_2(x,y)\left(\frac{dy}{dx}\right)^2+3a_3(x,y)\frac{dy}{dx}+
a_4(x,y)=0,
\]
with arbitrary coefficients $a_i(x,y)$ are form-invariant under
the change of the coordinates $$ x=f(u,v),\quad y=h(u,v)
 $$
and are equivalent to the system
\[
\frac{d^2
x}{ds^2}-a_3(x,y)\left(\frac{dx}{ds}\right)^2-2a_2(x,y)\left(\frac{dx}{ds}\right)\left(\frac{dy}{ds}\right)-
a_1(x,y)\left(\frac{dy}{ds}\right)^2=0,
\]
\[
\frac{d^2
y}{ds^2}+a_4(x,y)\left(\frac{dx}{ds}\right)^2+2a_3(x,y)\left(\frac{dx}{ds}\right)\left(\frac{dy}{ds}\right)+
a_2(x,y)\left(\frac{dy}{ds}\right)^2=0,
\]
having the form of geodesics of two dimensional affinely connected
space (
   with this aim the formulae
\[
\frac{d^2 y}{dx^2}=\frac{\dot x \ddot y-\dot y \ddot x}{(\dot
x)^3})
 \]
was used.

\section{From the  affinely connected space to the
Riemann space}

     Now we shall construct the Riemann space starting from a
     given affinely connected space defined by the second order
     ODE's.

      With this aim we use the notion of the Riemann extension of nonriemannian  space which was used
      earlier in the articles of author.

     Remind the basic properties of this construction.

     With help of the coefficients of affine connection of a given n-dimensional space
      can be introduced  2n-dimensional
     Riemann space $D^{2n}$ in local coordinates $(x^i,\Psi_i)$ having the metric of form
\begin{equation} \label{dryuma:eq11}
{^{2n}}ds^2=-2\Pi^k_{ij}(x^l)\Psi_k dx^i dx^j+2d \Psi_k dx^k
\end{equation}
\noindent where $\Psi_{k}$ are an additional coordinates.

  The important property of such type metric is that the geodesic
 equations of metric (\ref{dryuma:eq11})  decomposes into two parts
\begin{equation} \label{dryuma:eq12}
\ddot x^k +\Pi^k_{ij}\dot x^i \dot x^j=0,
\end{equation}
and
\begin{equation} \label{dryuma:eq13}
\frac{\delta^2 \Psi_k}{ds^2}+R^l_{kji}\dot x^j \dot x^i \Psi_l=0,
\end{equation}
where
\[
\frac{\delta \Psi_k}{ds}=\frac{d
\Psi_k}{ds}-\Pi^l_{jk}\Psi_l\frac{d x^j}{ds}
\]
and $R^l_{kji}$ are the curvature tensor of n-dimensional space
with a given affine connection.

 The first part (\ref{dryuma:eq12}) of the full system
is the system of equations for geodesic of basic space with local
coordinates $x^i$ and it do not contains the supplementary
coordinates $\Psi_k$.

 The second part (\ref{dryuma:eq13}) of the system  has the form
of linear $N\times N$ matrix system of second order ODE's for
supplementary  coordinates $\Psi_k$
\begin{equation} \label{dryuma:eq14}
\frac{d^2 \vec \Psi}{ds^2}+A(s)\frac{d \vec \Psi}{ds}+B(s)\vec
\Psi=0.
\end{equation}

   Remark that the full system of geodesics has the first integral
\begin{equation} \label{dryuma:eq15}
-2\Pi^k_{ij}(x^l)\Psi_k \frac{dx^i}{ds}\frac{dx^j}{ds}+2\frac{d
\Psi_k}{ds}\frac{dx^k}{ds}=\nu
\end{equation}
which is equivalent to the relation
\begin{equation} \label{dryuma:eq16}
2\Psi_k\frac{dx^k}{ds}=\nu s+\mu
\end{equation}
where $\mu, \nu$ are parameters.

   The geometry of extended space
connects with geometry of basic space. For example the property of
the space to be Ricci-flat $R_{ij}=0$ or symmetrical
$R_{ijkl;m}=0$ keeps also for the extended space.

    It is important to note that for extended  space having the metric (\ref{dryuma:eq11})
    all scalar curvature invariants are vanished.

    As consequence the properties of linear system of
equation (\ref{dryuma:eq13}-\ref{dryuma:eq14}) depending from the
the invariants of  $N\times N$ matrix-function
\[
E=B-\frac{1}{2}\frac{d A}{ds}-\frac{1}{4}A^2
\]
under change of the coordinates $\Psi_k$ can be of used for that.

   First applications the notion of extended spaces for the studying
of nonlinear second order ODE's connected with nonlinear dynamical
systems have been considered by author (V.Dryuma 2000-2008).

\section{Rigorous approach to geometry of  planar
systems}

       The system of paths of two-dimensional
     space $S_2$ in general form looks as

$$ \ddot x +\Pi^1_{11}(\dot
x)^2+2\Pi^1_{12}\dot x \dot y+\Pi^1_{22}(\dot y)^2=0, $$ $$ \ddot
y +\Pi^2_{11}(\dot x)^2+2\Pi^2_{12}\dot x \dot y+\Pi^2_{22}(\dot
y)^2=0, $$ where the coefficients $\Pi^k_{ij}=\Pi^k_{ji}$.

    The Riemann extension of the space $S_2$ is determined by the metric
$$  {^4}ds^2=-2 z \Pi^1_{11}d
x^2-2 t \Pi^2_{11}d x^2-4 z \Pi^1_{12}d x d y-4 t \Pi^2_{12}dx d
y-$$$$-2 z \Pi^1_{22}d y^2-2 t \Pi^2_{22}d y^2+2dx dz+2dy dt.
$$

       We shall apply such system of equations for the studying of the first order
      planar system of ODE's.

      With this aim we use the facts about the geometries of paths for which the equations
      of the paths admits independent linear first integrals.

$L.P.Eisenhart,\quad 1925$

A necessary condition to geodesic admit the linear first integral
$$
a_i(x,y)\frac{d x^i}{d s}=const
$$ is
$$
 a_{i;j}+a_{j;i}=0,
$$
 where
$$
 a_{i;j}=\frac{\partial a_i}{\partial x^j}-a_k \Gamma^k_{ij},
$$
 and  $\Gamma^k_{ij}$ are the Christoffel symbols of the
metric.

    We apply this conditions and their consequence
$$
 a_{i;j;k}+R^{m}_{k i j}a_m=0
$$
 where $R^i_{j k l}$ is the curvature tensor of the space to
determination of the coefficients of equations
    $\Pi^k_{ij}$ using the  vector $a_i$ in
    form
$$
a_i=[Q(x,y),-P(x,y),0,0].
$$

   By this  means that the first order of equation
$$
 \frac{d y}{d x}=\frac{Q(x,y)}{P(x,y)}
$$
 or
$$
 Q(x,y)d x-P(x,y)d y=0
$$ is an integral of the paths equations.

     With the help of these  conditions it is possible to state only three
     coefficients of affine connections $\Pi^k_{ij}$.

     For determination of others coefficients
we use yet another the first order equation
$$
\frac{d y}{dx}=-\frac{y(y-1)}{x(x-1)} ,
$$
 with a first integral
$$
y(x)=\frac{C(x-1)}{x-C}.
$$

   In this case a second vector $b_i$ is in form
$$
b_i=[y(y-1),x(x-1),0,0].
$$

    The equation  plays  an important role in
theory of
   of  planar the first order system of equations and was used in a famous article of
   Petrovsky-Landis (1956) (\cite{dryuma4:dryuma}).

     So, from the conditions on the metrics to admits  two linear first integrals
     the coefficients $\Pi^k_{ij}$ of the paths equation are uniquely determined and have the form
\[
\Pi^1_{11}={\frac {\left ({\frac {\partial }{\partial
x}}Q(x,y)\right )x\left (x- 1\right
)}{{y}^{2}P(x,y)-yP(x,y)+Q(x,y){x}^{2}-Q(x,y)x}}
\]
\[
\Pi^1_{22}=-{\frac {\left ({\frac {\partial }{\partial
y}}P(x,y)\right )x\left (x -1\right )}{{y}^{2}P(x,y)-y
P(x,y)+Q(x,y){x}^{2}-Q(x,y)x}}
\]
and corresponding expressions for the coefficients $\Pi^2_{11},
\Pi^2_{12},\Pi^2_{22}$

   Remark that last two equations are reduced at the independent equations
\[
\frac{ d^2 z}{ds^2}+M(s)\frac{d z}{ds}+N(s)z(s)+F(s)=0
\]
and
\[
\frac{ d^2 t}{ds^2}+U(s)\frac{d t}{ds}+V(s)t(s)+H(s)=0
\]
with the help of the first integral of geodesics
\[
z(s)\frac{ d x}{ds}+t(s)\frac{d y}{ds}-\alpha\frac{s}{2}-\beta=0
\]
of the metric (\ref{dryuma:eq16}).

     It is interested to note that such type of non homogeneous linear second order ODE's
      are connected with theory of first order systems of ODE's as the equations
      on the periods of corresponding  Abel integrals.

\section{The second order ODE's cubic  on the first
derivative }

     The first two equations of geodesic of the metric are
     equivalent to the one second order differential equation
\[
{\frac {d^{2}}{d{x}^{2}}}y(x)+{\frac {\left (\left ({\frac
{\partial } {\partial y}}P(x,y)\right ){x}^{2}-\left ({\frac
{\partial }{\partial y}}P(x,y)\right )x\right )\left ({\frac
{d}{dx}}y(x)\right )^{3}}{{y}^
{2}P(x,y)-yP(x,y)+Q(x,y){x}^{2}-Q(x,y)x}}\!+\!\]\[\!+\!{\frac
{\left (\left (P_x\!-\!Q_y\right ){x}^{2}\!+\!\left
(Q_y\!-\!P_x\!-\!2\,P\right )x\!+\!\left (P_y\right
){y}^{2}\!+\!\left (\!-\!2\,P\!-\!P_y\right )y\!+\!2\,P\right
)\left ( {\frac {d}{dx}}y\right
)^{2}}{{y}^{2}P\!-\!yP\!+\!Q{x}^{2}\!-\!Q x}}\!+\!\]\[+{\frac
{\left (\!-\!\left (Q_x \right ){x}^{2}\!+\!\left (Q_x\!+\!2\,Q
\right )x\!+\!\left (P_x\!-\!Q_y\right ){y}^{2}\!+\!\left
(2\,Q\!-\!P_x\!+\!Q_y \right )y\!-\!2\,Q\right ){\frac
{d}{dx}}y}{{y}^{2}P-yP+ Q{x}^{2}-Qx}}\!+\!\]\[+{\frac {-\left
({\frac {\partial }{\partial x} }Q(x,y)\right ){y}^{2}+\left
({\frac {\partial }{\partial x}}Q(x,y) \right
)y}{{y}^{2}P(x,y)-yP(x,y)+Q(x,y){x}^{2}-Q(x,y)x}}=0.  \qquad(14)
\]

    The  equation $(14)$ is of the form
\[
\frac{d^2
y}{dx^2}+a_1(x,y)\left(\frac{dy}{dx}\right)^3+3a_2(x,y)\left(\frac{dy}{dx}\right)^2+3a_3(x,y)\frac{dy}{dx}+
a_4(x,y)=0,
\]
and it has the invariants depending from the coefficients
$a_i(x,y)$  (R.Liouville, 1880, T.Tresse, 1886) under
transformations of variables $(x,y)$.

   As it was shown by (E.Cartan, 1924) these invariants in general case are same with the invariants
   of two-dimensional surface $V^2(x,y)$ in a four-dimensional projective space
   $RP^4(\xi^i)$ (under the reparametrization and the change of coordinates $\xi^i$).

     For example the invariants of equations with condition on coefficients
$ \nu_5=0 $ same with the invariants of developing surfaces in a
three-dimensional projective space $RP^3$.

   The invariants of R.Liouville have been used successful in theory of
ODE's and their applications in works author (V.Dryuma, 1984-). In
particular the second order ODE with a  Painleve property have the
condition  $\nu_5=0$.

    In our case the equation has the particular integral
\[
\frac{d y}{d x}=\frac{Q(x,y)}{P(x,y)}
\]
 and the function
$$
 y(x)=\frac{C(x-1)}{x-C}.
$$  as the first integral.

     From geometrical point of view the equation $(14)$ corresponds two-dimensional surface in the $RP^4$-space.

     In this context it is interested to note the relation with the Petrovsky-Landis
     theory of limit cycles of the equation
\[
\frac{dy}{dx}=\frac{a_0+a_1x+a_2y+a_{11}x^2+a_{12}xy+a_{22}y^2}{b_0+b_1x+b_2y+b_{11}x^2+b_{12}xy+b_{22}y^2}.
\]

   In the famous article of $P-L$
was developed approach to the studying of the problem of the limit
cycles of the first order quadratic equation.

  Let us recall the basic facts of the Petrovsky-Landis theory.

   For  studying of the closed curves  of quadratic the first order equation
   is considered  the equation
\[
 \frac{d y}{dx}=-\frac{y(y-1)}{x(x-1)}
\]
with solution
\[
 y(x)=\frac{C(x-1)}{(x-C)}.
\]

    As it was showed for the closed curves the parameter $C$ satisfies the algebraic
equations
\[
\sum a_n(\mu_i)C^n=0
\]
where the coefficients $a_n(\mu_i)$ are dependent from the
parameters of the quadratic equation.

    Such type of equation arises  from the  condition

\[
\int_{c}{\frac{(x-C)^2[x(x-1)P(x,y)+y(y-1)Q(x,y)]}{x^3(x-1)^3}}d
x=0,
\]

 which lead to determination of the closed integral curves
of the first order system of equations
\[
\frac{dy}{dx}=\frac{a_0+a_1x+a_2y+a_{11}x^2+a_{12}xy+a_{22}y^2}{b_0+b_1x+b_2y+b_{11}x^2+b_{12}xy+b_{22}y^2}.
\]

    As it was shown by Petrovsky-Landis in this case the
parameter $C$ is determined from calculations of the residues of
integral after substitution of the expression
\[
y=\frac{C(x-1)}{(x-C)}
\]
in it.

   The simplest of them are

\[
\left ({\it b2}+2\,{\it b0}+{\it a12}+{\it a1}+{\it a2}+2\,{\it
a0}+{ \it b12}+{\it b1}\right ){C}^{2}+\]\[+\left (-2\,{\it
a0}-{\it b12}-{\it a2 }-{\it b1}+2\,{\it b22}-2\,{\it b0}\right
)C-2\,{\it b22}-{\it b2}=0,  \quad x=0,
\]
\[
\left (-{\it b2}-2\,{\it b0}-{\it a12}-{\it a1}-{\it a2}-2\,{\it
a0}-{ \it b12}-{\it b1}\right ){C}^{2}+\]\[+\left ({\it a2}+{\it
b1}+{\it a12}+2 \,{\it b0}+2\,{\it a0}-2\,{\it a11}\right
)C+2\,{\it a11}+{\it a1}=0, \quad  x=1,
\]

\[\left ({\it b12}-2\,{\it b22}\right )C+{\it b2}+2\,{\it b22}=0, \quad  x=C.
\]

\[
\left (-{\it a12}+2\,{\it a11}\right )C-2\,{\it a11}-{\it a1}=0,
\quad   x=\infty.
\]

   Others equations on $C$ from the combinations of these
   conditions are followed.

     According the results of $(P-L)$ general quantity of the
     values $C$ defined by a such type of equations is equal $14$ and this
     number coincides with the quantity of closed solutions determined by the
     quadratic equation.

      On other side $11$ curves from $14$ can be transformed into the small neighborhood
      of the essential singular points of the equation
      \[
 \frac{d y}{dx}=-\frac{y(y-1)}{x(x-1)}.
\]

They are: $(0,1)$, $(1,0)$, $(0,0)$, $(1,1)$ and so on...

     As result only three closed curves do not be transformed into the neighborhood of the singular
      points and so the quantity of
     the limit cycles defined by the equation is equal three.

       It is interested to note that a some conditions on the
       parameter $C$ are appeared in context of the second order
       ODE.

        In fact the result of joint consideration of the first order equations
$$
y'=\frac{Q(x,y)}{P(x,y)},\quad y'=\frac{y(y-1)}{x(x-1)}
$$
the function
$$
y(x)=\frac{C(x-1)}{(x-C)}
$$
and the second order ODE $(13)$ we get the conditions

$$
\alpha(x,y)C^5+\beta(x,y)C^4+\gamma(x,y)C^3+\delta(x,y)C^2+\epsilon(x,y)C+\mu(x,y)=0,
$$
 where
\[
\alpha(x,y)=\left ({\it a12}+{\it b12}\right ){y}^{2}+\]\[+\left
({\it b1}+{\it b2}+ \left (2\,{\it a22}+2\,{\it a11}+2\,{\it
b22}+2\,{\it b11}\right )x+{ \it a2}+{\it a1}\right )y+\]\[+\left
({\it a12}+{\it b12}\right ){x}^{2}+ \left ({\it a1}+{\it b2}+{\it
b1}+{\it a2}\right )x+2\,{\it a0}+2\,{ \it b0}  ,
\]
\[
\beta(x,y)=2\,{\it b22}\,{y}^{3}+\left (\left (-5\,{\it a12}-{\it
b12}-2\,{\it b22}\right )x+2\,{\it b2}-{\it b12}\right
){y}^{2}+\]\[+\left (\left (\!-\!4\,{ \it a22}\!-\!3\,{\it
b2}\!-\!5\,{\it a1}\!-\!3\,{\it a2}\!-\!2\,{\it b11}\!-\!6\,{\it
b22}\!-\!{\it b1}\right )x\!-\!{\it a2}\right)+\]\[+\left(\!\left
(\!-\!4\,{\it b11}\!+\!{\it a12}\!-\!10\,{\it a11}\!-\!{\it
b12}\!-\!4\,{\it a22}\right ){x}^{2}\!+\!2\,{\it b0}\!-\!2\,{\it
b2}\!-\!{ \it b1}\right )y\!+\]\[+\!\left (\!-\!2\,{\it
a12}\!+\!2\,{\it a11}\right ){x}^{3}\!+\! \left (\!-\!2\,{\it
a12}\!-\!2\,{\it b1}\!-\!2\,{\it a1}\!-\!3\,{\it b12}\!-\!2\,{\it
a2} \right ){x}^{2}\!-\!4\,{\it b0}\!+\]\[+\!\left (\!-\!6\,{\it
a0}\!-\!2\,{\it b1}\!-\!4\,{\it b0 }\!-\!3\,{\it b2}\!-\!{\it
a1}\!-\!2\,{\it a2}\right )x\!-\!2\,{\it a0} ,
\]
\[
\mu(x,y)=-{y}^{2}{x}^{5}{\it a12}+\left (-2\,{x}^{6}{\it
a11}+\left ({\it a12}- {\it a1}\right ){x}^{5}\right
)y+2\,{x}^{6}{\it a11}+{x}^{5}{\it a1}.
\]

    From these relation we get the relations between the parameter $C$
    and coefficients $a_{ij},~b_{ij},~a_i,~b_i$

    As example at the values $x=0$ and $x=1$  we get conditions on the value $C$
\[ \left ({\it a2}+{\it a12}+{\it b1}+{\it b12}+{\it b2}+2 \,{\it
a0}+{\it a1}+2\,{\it b0}\right ){C}^{2}+\]\[+\left (-2\,{\it
b0}-2\, {\it a0}-{\it b1}-{\it a2}-{\it b12}+2\,{\it b22}\right
)C-{\it b2}-2 \,{\it b22}=0, \] \[
 \left ({\it a2}+{\it
a12}+{\it b1}+{\it b12}+{\it b2}+2\,{\it a0}+{\it a1}+2\,{\it
b0}\right ){C}^{2}+\]\[+\left (2 \,{\it a11}-2\,{\it b0}-{\it
b1}-2\,{\it a0}-{\it a12}-{\it a2}\right )C-2\,{\it a11}-{\it
a1}=0. \]

   The substitution $(x=C, y=1-C)$  lead
   to the conditions on the value $C$
$$
\left ({\it b12}-2\,{\it b22}\right )C+2\,{\it b22}+{\it b2}=0.
$$

   After substitution $(y=1-x,x=1, C=1/C1)$ we get
\[
 \left ({\it a1}+2\,{\it
a11}\right ){{\it C1}}^{2}+\left ({\it a12}-2 \,{\it a11}+{\it
b1}+{\it a2}+2\,{\it a0}+2\,{\it b0}\right ){\it C1}-\]\[- {\it
a1}-{\it a2}-{\it a12}-2\,{\it b0}-{\it b12}-{\it b1}-{\it b2}-2
\,{\it a0}=0, \]
  and the substitution $(y=1-x,x=0, C=1/C1)$
lead to the condition \[
 \left (-2\,{\it b22}-{\it b2}\right
){{\it C1}}^{2}+\left (-2\,{\it b0 }-{\it b12}-{\it b1}-{\it
a2}-2\,{\it a0}+2\,{\it b22}\right ){\it C1} +\]\[+{\it a1}+{\it
a2}+{\it a12}+2\,{\it b0}+{\it b12}+{\it b1}+{\it b2}+2 \,{\it a0}
=0.
 \]

   In result of a such type consideration we have got the conditions
of Petrovskii-Landis article on parameter $C$.

\section{Three dimensional homogeneous system}

    As it was shown in article (V.Dryuma,2006) between the planar system
$$
 \frac{d x}{dt}=a_0+a_1x+a_2y+a_{11}x^2+a_{12}xy+a_{22}y^2,$$
 $$
  \frac{d y}{dt}=b_0+b_1x+b_2y+b_{11}x^2+b_{12}xy+b_{22}y^2,
$$

and a spatial homogeneous quadratic system of equations
$$ \frac{d x}{dt}=P(x,y,z),\quad \frac{d y}{dt}=Q(x,y,z),
\quad \frac{d z}{dt}=Q(x,y,z)
 $$
 of the form
$$
 \frac{d x}{dt}=4a_0 z^2\!+\!4a_2 yz\!+\!(3a_1-b_2)xz\!+\!4 a_{22} y^2\!+\!(3a_{12}-2b_{22})xy\!+
 \!(2a_{11}-b_{12})x^2,
 $$
 $$
  \frac{dy}{dt}=4b_0z^2\!+\!4b_1 xz\!+\!(3b_2-a_1)yz+4b_{11}x^2\!+\!(3b_{12}-2a_{11})xy\!+
  \!(2b_{22}-a_{12})y^2,$$
  $$
  \frac{dz}{dt}=-(a_1+b_2)z^2-(2b_{22}+a_{12})yz-(2a_{11}+b_{12})xz
 $$
exists some connections.

    For a such system the condition
    $$
    \frac{\partial P(x,y,z)}{\partial x}+\frac{\partial Q(x,y,z)}{\partial
    y}+\frac{\partial R(x,y,z)}{\partial z}=0
$$
is fulfilled and in variables
$$
    \xi(t)=\frac{x(t)}{z(t)},\quad \eta(t)=\frac{y(t)}{z(t)}
$$
it takes the form of the equation
 $$
 \frac{d \xi}{d \eta}=\frac{a_0+a_1\xi+a_2\eta+a_{11}\xi^2+a_{12}\xi \eta+a_{22}\eta^2}
 {b_0+b_1\xi+b_2\eta+b_{11}\xi^2+b_{12}\xi \eta+b_{22}\eta^2}
 $$
equivalent the planar system .

    It is significant that the spatial system  in the variables
$$
x(t)=\frac{d X(t)}{d t}, \quad y(t)=\frac{d Y(t)}{d t},\quad
z(t)=\frac{d Z(t)}{d t}
$$ takes the
form of geodesic equations of a three dimensional space
   \begin{equation}\label{dr:eq5}
\frac{d^2 X^i}{d t^2}+\Gamma^i_{jk}\frac{d X^j}{d t}\frac{d X^k}{d
t}=0
\end{equation}
in local coordinates $X^i=(X(t),~Y(t),~Z(t)$.

 In doing so the coefficient $\Gamma^i_{jk}$ are
constant and depend on the parameters $a_i,~a_{ij},b_i,~b_{ij}$.

The set of such coefficients can be considered as the coefficients
of affine connection on a three dimensional space $H_3$ in local
coordinates $X^i$.

    From geometrical point of view through the equations ~(\ref{dryuma:eq5}) on the space $H_3$
    the structure of affinely connected space with constant coefficients of connection $\Gamma^i_{jk}$
    is determined.

\section{Six-dimensional Riemann space}

 For the $n$-dimensional space equipped with affine connection $\Gamma^i_{jk}$  the
 metrics of the Riemann extension $V_{2n}$  has the form
\begin{equation}\label{dreq6}
^{2n}ds^2=-2\Gamma^i_{jk}dX^j d X^k d \xi_i-2d \xi_i dX^i
\end{equation}
where $\chi_i$ are  an additional coordinates.

     Non zero coefficients of affine connections of the space $H_3$ are
$$
 \Gamma^1_{33}=-4a_0,\quad
\Gamma^1_{23}=-2a_2,\quad
\Gamma^1_{13}=\frac{1}{2}(b_2-3a_1),\quad
\Gamma^1_{22}=-4a_{22},$$$$
\Gamma^1_{12}=\frac{1}{2}(2b_{22}-3a_{12}),\quad
\Gamma^1_{11}=b_{12}-2a_{11}, $$ $$
 \Gamma^2_{33}=-4b_0,\quad
\Gamma^2_{13}=-2b_1,\quad
\Gamma^2_{23}=\frac{1}{2}(a_1-3b_2),\quad
\Gamma^2_{11}=-4b_{11},$$$$
\Gamma^2_{12}=\frac{1}{2}(2a_{11}-3b_{12}),\quad
\Gamma^2_{22}=a_{12}-2b_{22}, $$ $$
 \Gamma^3_{33}=a_1+b_2,\quad
\Gamma^3_{23}=\frac{1}{2}(a_{12}+2b_{22}),\quad
\Gamma^3_{13}=\frac{1}{2}(2a_{11}+b_{12}).
 $$

    According with the ~(\ref{dreq6}) the Riemann metric of six dimensional extended
     space $V_6$ has the form
$$
 ^{6}{{\it ds}}^{2}=\left (-2\,U{\it b12}+4\,U{\it a11}+8\,{\it b11}\,V
\right ){{\it dx}}^{2}+\left (8\,{\it a22}\,U-2\,V{\it
a12}+4\,V{\it b22}\right ){{\it dy}}^{2}+$$$$+\left (8\,{\it
b0}\,V-2\,W{\it b2}+8\,{\it a0}\,U-2\,W{\it a1}\right ){{\it
dz}}^{2}+\left (-4\,U{\it b22}-4\,V {\it a11}+6\,V{\it
b12}+6\,U{\it a12}\right ){\it dx}\,{\it dy}+$$$$+\left (6\,U{\it
a1}-2\,U{\it b2}-4\,W{\it a11}-2\,W{\it b12}+8\,{\it b1}\,V \right
){\it dx}\,{\it dz}+$$$$+\left (-2\,W{\it a12}+8\,{\it
a2}\,U+6\,V{ \it b2}-2\,V{\it a1}-4\,W{\it b22}\right ){\it
dy}\,{\it dz}+2\,{\it dx}\,{\it dU}+2\,{\it dy}\,{\it dV}+2\,{\it
dz}\,{\it dW} . $$

   Remark that we use the denotes $(x,~y,~z)$ for the coordinates $(X,~Y,~Z)$.

        Geodesic of the metric consist from  two parts.
 Nonlinear system of coupled equations on coordinates $(x,y,z)$
$$ {\frac {d^{2}}{d{s}^{2}}}x(s)+\left (-2\,{\it a11}+{\it
b12}\right ) \left ({\frac {d}{ds}}x(s)\right )^{2}+2\,\left
(-3/2\,{\it a12}+{\it b22}\right )\left ({\frac {d}{ds}}x(s)\right
){\frac {d}{ds}}y(s)+$$$$+2\, \left (-3/2\,{\it a1}+1/2\,{\it
b2}\right )\left ({\frac {d}{ds}}x(s) \right ){\frac
{d}{ds}}z(s)-4\,{\it a22}\,\left ({\frac {d}{ds}}y(s) \right
)^{2}-4\,{\it a2}\,\left ({\frac {d}{ds}}y(s)\right ){\frac {d}
{ds}}z(s)-4\,{\it a0}\,\left ({\frac {d}{ds}}z(s)\right )^{2}=0,
$$ $${\frac {d^{2}}{d{s}^{2}}}y(s)-4\,{\it b11}\,\left ({\frac
{d}{ds}}x(s) \right )^{2}+2\,\left (-3/2\,{\it b12}+{\it
a11}\right )\left ({\frac {d}{ds}}x(s)\right ){\frac
{d}{ds}}y(s)-4\,{\it b1}\,\left ({\frac {d} {ds}}x(s)\right
){\frac {d}{ds}}z(s)+\]\[+\left (-2\,{\it b22}+{\it a12} \right
)\left ({\frac {d}{ds}}y(s)\right )^{2}+2\,\left (-3/2\,{\it b2
}+1/2\,{\it a1}\right )\left ({\frac {d}{ds}}y(s)\right ){\frac
{d}{ds }}z(s)-4\,{\it b0}\,\left ({\frac {d}{ds}}z(s)\right )^{2}
=0,
 $$
$$ {\frac {d^{2}}{d{s}^{2}}}z(s)+2\,\left ({\it a11}+1/2\,{\it
b12} \right )\left ({\frac {d}{ds}}x(s)\right ){\frac
{d}{ds}}z(s)+2\, \left (1/2\,{\it a12}+{\it b22}\right )\left
({\frac {d}{ds}}y(s) \right ){\frac {d}{ds}}z(s)+\left ({\it
a1}+{\it b2}\right )\left ({ \frac {d}{ds}}z(s)\right )^{2}  =0.
$$

   And the linear system of equations on coordinates $(U,V,W)$
$$ {\frac {d^{2}}{d{s}^{2}}}U(s)+{\it A1}\,{\frac
{d}{ds}}U(s)+{\it B1}\, {\frac {d}{ds}}V(s)+{\it C1}\,{\frac
{d}{ds}}W(s)+{\it E1}\,U(s)+{\it F1}\,V(s)+{\it H1}\,W(s) =0 $$
 $$
{\frac {d^{2}}{d{s}^{2}}}V(s)+{\it A2}\,{\frac {d}{ds}}U(s)+{\it
B2}\, {\frac {d}{ds}}V(s)+{\it C2}\,{\frac {d}{ds}}W(s)+{\it
E2}\,U(s)+{\it F2}\,V(s)+{\it H2}\,W(s) =0 $$
 $$
 {\frac {d^{2}}{d{s}^{2}}}W(s)+{\it A3}\,{\frac {d}{ds}}U(s)+{\it B3}\,
{\frac {d}{ds}}V(s)+{\it C3}\,{\frac {d}{ds}}W(s)+{\it
E3}\,U(s)+{\it F3}\,V(s)+{\it H3}\,W(s) =0, $$
 where
the coefficients $(Ai,~Bi,~Ci)$ are depended from the parameters
$(a,b)$ and derivatives $(\dot x,~\dot y,~\dot z)$ with respect to
the parameter $s$.

    Taking in consideration $x,y,~z$-equations the
    $U,~V,~W$- equations can be one time integrated and take
     the form
$$ {\frac {d}{ds}}U(s)=\left
(2\,{\it a11}+{\it b12}\right )z(s)W(s)+$$$$+\left (\left
(-3\,{\it b12}+2\,{\it a11 }\right )y(s)-8\,{\it b11}x(s)-4\,{ \it
b1}z(s)\right)V(s)+$$$$+\left (\left (2\,{\it b12}-4 \,{\it
a11}\right )x(s)+\left (-3\,{\it a12}+2\,{\it b22}\right
)y(s)+\left ({\it b2}-3\,{\it a1}\right )z(s)\right )U(s),
 $$
$$
{\frac {d}{ds}}V(s)=\left ({\it a12}+2\,{\it b22}\right
)z(s)W(s)+$$$$+\left (\left (-3\,{\it a12}+2\,{\it b22 }\right
)x(s)-8\,{\it a22}y(s)-4\,{ \it a2}z(s)\right )U(s)+$$$$+\left
(\left (-3\,{\it b12}+2 \,{\it a11}\right )x(s)+\left (2\,{\it
a12}-4\,{\it b22 }\right )y(s)\!+\!\left(\!-\!3\,{\it b2}\!+\!{\it
a1}\right )z(s)\right)V(s),
 $$
$$
{\frac {d}{ds}}W(s)=\left (\left ({\it b2}-3\,{\it a1}\right
)x(s)-8\,{\it a0}z(s)-4{\it a2}\,y(s)\right )U(s)+$$$$+\left
(\left (-3\,{\it b2}+{\it a1}\right )y(s)-8\,{\it b0}z(s)-4\,{\it
b1}x(s)\right )V(s)+$$$$+\left (\left (2\,{\it a11}+{\it
b12}\right )x(s)+\left ({\it a12}+2\,{\it b22}\right
)y(s)+2\,\left ({\it a1}+{\it b2}\right )z(s)\right )W( s).
$$

   In result we have got a six-dimensional Riemann space associated with
   a quadratic the first order system of equations.

   An investigation  the properties of the metric at the change of parameters $a,b$ may be useful for the theory
of a such type of the systems.

   In particular a study of the Killing properties of the metric
   allow us to get information on particular integrals of geodesic equations.

\section{On the surfaces defined by spatial system of
equations}

The equation of  surfaces $z=z(x,y)$ defined by the spatial system
of equations
   ~(\ref{dryuma:eq4}) has the form of the first order p.d.e
\begin{equation}\label{dryuma:eq161}
 z_x
 \left( 4\,{\it a0}\, \left( z   \right) ^{2}+
 \left( 4\,{\it a2}\,y+ \left( 3\,{\it a1}-{\it b2} \right) x \right)
z  +4\,{\it a22}\,{y}^{2}\right)+z_x \left( \left( 3\,{\it
a12}-2\,{ \it b22} \right) xy+ \left( 2\,{\it a11}-{\it b12}
\right) {x}^{2}
 \right) + \]\[+z_y \left(
  4 b0 z^2+ \left( 3\,{\it b2}-{\it a1} \right) y z+4\,{\it b11}\,{x
}^{2}+4\,{\it b1}\,x
  z \right)+z_y \left(\left( 2\,{\it b22}-{\it a12} \right)
{y}^{2}+ \left( -2\,{\it a11}+3\,{\it b12} \right) xy \right) +
\]\[+\left( {\it b2}+{\it a1} \right) \left( z   \right) ^{2}+
\left(  \left( 2\,{\it b22}+{\it a12} \right) y+ \left( 2\,{\it
a11}+{\it b12} \right) x \right) z.
\end{equation}

     An examples of solutions of this equation  were obtained
     by the method of $(u,v)$-transformation developed
      earlier by author.

   To integrate the partial nonlinear first order differential equation
\begin{equation}\label{dryuma:eq16}
F(x,y,z(x,y),z_x,z_y)=0
 \end{equation} we use a following change
of the functions and  variables
\begin{equation}\label{dryuma:eq17}
z(x,y)\rightarrow u(x,t),\quad y\rightarrow v(x,t),\quad
z_x\rightarrow u_x-\frac{v_x}{v_t}u_t,\quad v_y \rightarrow
\frac{u_t}{v_t}.
\end{equation}
  In result instead of the equation (\ref{dryuma:eq16}) one get the
  relation between the new variables ~$u(x,t)$ and $v(x,t)$ and
  their partial derivatives
\begin{equation}\label{dryuma:eq18}
\Phi(u,v,u_x,u_t,v_x,v_t)=0.
  \end{equation}

    In many cases the solution of last equation is a more simple problem
    than solution of the equation
    (\ref{dryuma:eq16}).

     In result of application of the $(u,v)$-transformation
     the equation for $z=z(x,y)$  takes the form of (\ref{dryuma:eq18}).

     Using the substitution
\[
u(x,t)=t{\frac {\partial }{\partial
t}}\omega(x,t)-\omega(x,t),\quad v(x,t)={\frac {\partial
}{\partial t}}\omega(x,t),
\]
where $
 \omega(x,t)=xA(t),
$ we find from a given relation the equation for the function
 $A(t)$
\[
 \left (-{t}^{2}{\it b2}-t{\it b22}+{\it a22}\,A(t)-{t}^{3}{\it b0}+{
\it a0}\,A(t){t}^{2}+{\it a2}t \,A(t)\right )\left ({\frac
{d}{dt}}A(t) \right )^{2}+\]\[+\left (-2\,{\it a0}\,\left
(A(t)\right )^{2}t-t{\it b12}- {\it a2}\,\left (A(t)\right
)^{2}-{t}^{2}{\it b1}+t{\it a1}\,A(t)\right){\frac {d} {dt}}A(t)
+\]\[+\left(2\,{ t}^{2}{\it b0}\,A(t)+t{\it b2}\,A(t)+{\it
a12}\,A(t)\right ){\frac {d} {dt}}A(t)-\]\[\!-\!t{\it
b11}\!-\!t{\it b0}\,\left (A(t)\right )^{2}\!-\!{\it a1}\, \left
(A(t)\right )^{2}\!+\!t{\it b1}\,A(t)\!+\!{\it a11}\,A(t)\!+\!{\it
a0}\, \left (A(t)\right )^{3}=0.
\]

    Solutions of this equation depend  from the parameters $a,~b$ and
     can play a key value in theory  of the quadratic systems.

     Another type of substitution
\[
v(x,t)=t{\frac {\partial }{\partial
t}}\omega(x,t)-\omega(x,t),\quad u(x,t)={\frac {\partial
}{\partial t}}\omega(x,t),
\]
where $
 \omega(x,t)=xB(t),
$ lead to the equation on the function $B(t)$
\begin{equation}\label{dryuma:eq191}
\left ({\it a22}\,B(t){t}^{2}+{\it a2}\,B(t)t+{\it a0}\,B(t)+{\it
b22} \,{t}^{2}+t{\it b2}+{\it b0}\right )\left ({\frac
{d}{dt}}B(t)\right ) ^{2}+\]\[+\left (t{\it b12}-{\it a2}\,\left
(B(t)\right )^{2}+{\it b1}+{ \it a1}\,B(t)-2\,{\it
b22}\,tB(t)-2\,{\it a22}\,\left (B(t)\right )^{2 }t+t{\it
a12}\,B(t)-{\it b2}\,B(t)\right ){\frac {d}{dt}}B(t)-\]\[-{\it a12
}\,\left (B(t)\right )^{2}-{\it b12}\,B(t)+{\it a11}\,B(t)+{\it
a22}\, \left (B(t)\right )^{3}+{\it b22}\,\left (B(t)\right
)^{2}+{\it b11}=0
 \end{equation}
 which is a more simple than previous and also can be
useful in theory of quadratic systems.

    Let us consider some examples.

    In the variables $$B(t)=S,~\frac{d B(t)}{ dt}=T$$ the equation
    (\ref{dryuma:eq191}) determines algebraic curve
    \[
    F(S,T,t,a,b)=0
    \]
having  genus $g=1$ or $g=0$ in depending on the parameters
$a,~b$.

     As example at the conditions
     \[
     b0=0,~a0=0,~b2=0,~a1=0,~a2=0,~a22=0,~b11=0,~a12=0,~a11=b12,~b1=0
     \]
the equation (\ref{dryuma:eq191}) takes the form
\[
{\it b22}\,{t}^{2}\left ({\frac {d}{dt}}A(t)\right )^{2}+\left
(-2\,{ \it b22}\,tA(t)+t{\it b12}\right ){\frac {d}{dt}}A(t)+{\it
b22}\, \left (A(t)\right )^{2}=0
\]
and defines algebraic curve of the genus $g=0$.

    Integral of this equation is defined by the relation
    \[
    \ln (t)+\ln (\sqrt {-4\,{\it b22}\,B(t){\it b12}+{{\it b12}}^{2}}-{
\it b12})-{\frac {{\it b12}}{\sqrt {-4\,{\it b22}\,B(t){\it
b12}+{{ \it b12}}^{2}}-{\it b12}}}-\]\[-\ln (\sqrt {-4\,{\it
b22}\,B(t){\it b12}+{{ \it b12}}^{2}}+{\it b12})-{\frac {{\it
b12}}{\sqrt {-4\,{\it b22}\,B(t ){\it b12}+{{\it b12}}^{2}}+{\it
b12}}}-1/2\,{\frac {{\it b12}}{{\it b22}\,B(t)}}-\ln (B(t))-\]\[-{\it
\_C1}=0.
\]

  After the inverse $(u,v)$- transformation  with the help of the function $B(t)$ can be find the solution
   of the equation (\ref{dryuma:eq161}) at indicated above the
   values of parameters.

\section{ Eight-dimensional Riemann space for the Lorenz
system of equations}

     To investigation of the properties of classical  Lorenz equations
 \begin{equation}\label{dryuma:eq19}
\frac{d x}{ds}=\sigma(y-x),\quad \frac{d y}{ds}=r x-y- x z, \quad
\frac{d z}{ds}=-b z+xy
\end{equation}
we use its presentation in the form
\[
{\frac {d}{ds}}\xi(s)=1/5\,\left (b-4\,\sigma+1\right )\xi\,\rho+
\sigma\,\eta\,\rho,
\]
\[
{\frac {d}{ds}}\eta(s)=-\xi\,\theta+r\xi\,\rho+1/5\,\left
(b+\sigma-4 \right )\eta\,\rho,
\]
\[
{\frac {d}{ds}}\theta(s)=\xi\,\eta+1/5\,\left (\sigma-4\,b+1\right
) \rho\,\theta ,
\]
\[
{\frac {d}{ds}}\rho(s)=1/5\,\left (\sigma+1+b\right ){\rho}^{2}.
\]

   The relation between both systems is defined by the conditions
\[
   x(s)=\frac{\xi}{\rho},\quad y(s)=\frac{\eta}{\rho},\quad
   z(s)=\frac{\theta}{\rho}.
\]

    Four dimensional system can be presented in the form
\[
    \frac{d^2 X^i}{d s^2} +\Gamma^i_{jk}\frac{d X^j}{d s}\frac{d X^k}{d
    s}=0,
\]
which allow us to consider it as geodesic equations of the space
with constant  affine connection.

    Nonzero components of connection are
    \[
    \Gamma^1_{14}=\frac{4\sigma-b-2}{10},\quad \Gamma^1_{24}=-\frac{\sigma}{2},
    \quad \Gamma^2_{13}=\frac{1}{2},\]\[ \Gamma^2_{14}=-\frac{r}{2},\quad
    \Gamma^2_{34}=\frac{4-\sigma-b}{10},\quad \Gamma^3_{34}=\frac{4b-\sigma-1}{10},
    \]\[ \Gamma^3_{12}=-\frac{1}{2},\quad\Gamma^4_{44}=-\frac{\sigma+b+1}{5}
    \]

     The metric of associated space is
      \begin{equation}\label{dryuma:eq20}
     ^{8}ds^2=2/5\,\left(b+\sigma+1 \right){{\it du}}^{2}V+\]\[+\left (2/5\,{\it dz}\,{\it du}+2\,{\it
dx} \,{\it dy}-8/5\,{\it dz}\,{\it du}\,b+2/5\,{\it dz}\,{\it
du}\,\sigma \right )U+\]\[+\left (2\,r{\it dx}\,{\it du}+2/5\,{\it
dz}\,{\it du}\, \sigma-2\,{\it dx}\,{\it dz}-8/5\,{\it dz}\,{\it
du}+2/5\,{\it dz}\,{ \it du}\,b\right )Q+\]\[+\left (-8/5\,{\it
dx}\,{\it du}\,\sigma+2\,\sigma \,{\it dy}\,{\it du}+2/5\,{\it
dx}\,{\it du}\,b+2/5\,{\it dx}\,{\it du }\right )P+\]\[+2\,{\it
dy}\,{\it dQ}+2\,{\it dz}\,{\it dU}+2\,{\it du}\,{ \it dV}+2\,{\it
dx}\,{\it dP}. \end{equation}

     After integration geodesic of additional coordinates take the form
     \[
{\frac {d}{dt}}P(t)-\left (z(t)-ru(t) \right
)Q(t)-\frac{1}{5}\left (-b-1+4 \sigma\,\right)u(t)P(t)+y(t)U(t)
=0,
\]
\[
{\frac {d}{dt}}Q(t)+x(t)U(t)+\sigma\,u(t)P(t)=0,
\]
\[
{\frac {d}{dt}}U(t)+\frac{1}{5}\left(\left( \sigma\,+b-4
\right)u(t)-x(t)\right )Q(t)+\frac{1}{5} \left
(-4b+1+\sigma\right)\,u(t)U(t)=0,
\]
\[
{\frac {d}{dt}}V(t)+1/5\,\left(b+\sigma+1\right)u(t)V(t)-\frac
{x(t)y(t)U(t)}{u (t)}+\frac {x(t)Q( t)z(t)}{u(t)}=0.
\]

     Properties of this system  from the parameters are dependent
     and can be investigated with the help of the Wilczynski invariants.

\section{Laplace operator}

   In theory of Riemann spaces the equation
 \begin{equation}\label{dryuma:eq21}
L\psi=g^{ij}(\frac{\partial ^2}{\partial x^i \partial x
^j}-\Gamma^k_{ij}\frac{\partial}{\partial x^k})\psi(x)=0
\end{equation}
can be used to the study of the properties of the space.

For the eight-dimensional space with metric (\ref{dryuma:eq20})
corresponded the Lorenz system we get the equation on the function
$\psi(x,y,z,u,P,Q,U,V)$
\[
-2\,U{\frac {\partial ^{2}}{\partial P\partial Q}}\psi+1/5\,\left
({\frac {\partial }{\partial V}}\psi \right )\sigma+2\,Q{\frac
{\partial ^{2}}{\partial P\partial U}}\psi+1/5\,\left ({\frac
{\partial }{\partial V}}\psi\right )b+8/5\,\left ({\frac {\partial
^{2}}{\partial P
\partial V}}\psi\right )P\sigma-\]\[-2/5\,\left ({\frac {
\partial ^{2}}{\partial P\partial V}}\psi\right )P-2
\,\sigma\,P{\frac {\partial ^{2}}{\partial Q\partial
V}}\psi-2\,\left ({\frac {\partial ^{2}}{\partial P\partial
V}}\psi\right )rQ+8/5\,\left ({\frac {\partial ^{2}}{\partial U
\partial V}}\psi\right )Q-2/5\,\left ({\frac {
\partial ^{2}}{\partial U\partial V}}\psi\right )U
\sigma+\]\[+8/5\,\left ({\frac {\partial ^{2}}{\partial U\partial
V}}\psi\right )Ub-2/5\,\left ({\frac {\partial ^{2}}{\partial
U\partial V}}\psi\right )U+2\,{\frac {\partial ^{2}}{
\partial P\partial x}}\psi-4/5\,{\frac {\partial }{
\partial V}}\psi+2\,{\frac {\partial ^{2}}{\partial Q
\partial y}}\psi+2\,{\frac {\partial ^{2}}{\partial U
\partial z}}\psi+2\,{\frac {\partial ^{2}}{\partial V
\partial u}}\psi-\]\[-2/5\,\left ({\frac {\partial ^{2}}{
\partial P\partial V}}\psi\right )Pb-2/5\,V\left ({
\frac {\partial ^{2}}{\partial {V}^{2}}}\psi\right )b -2/5\,V\left
({\frac {\partial ^{2}}{\partial {V}^{2}}}\psi\right
)\sigma-2/5\,V{\frac {\partial ^{2}}{\partial {V}^{2}}}
\psi-2/5\,\left ({\frac {\partial ^{2}}{\partial U
\partial V}}\psi\right )Qb-\]\[-2/5\,\left ({\frac {
\partial ^{2}}{\partial U\partial V}}\psi\right )Q
\sigma=0.
\]

   This equation has varies type of particular solutions.

   As example
\[
\psi(P,Q,U,V)=H(P,Q,U)+VP
\]
where the function $\psi(P,Q,U,V)$ satisfies the equation
   \[
   2\,Q{\frac {\partial ^{2}}{\partial P\partial U}}H(P,Q,U)+9/5\,P\sigma
-2\,U{\frac {\partial ^{2}}{\partial P\partial
Q}}H(P,Q,U)-1/5\,Pb-2\, rQ-6/5\,P=0.
\]

   Its solution is in form
   \[
H(P,Q,U)=\left (-3/10-1/20\,b+{\frac {9}{20}}\,\sigma\right
)\arctan({ \frac {Q}{U}}){P}^{2}+PrU+{\it \_F1}(Q,U)+{\it
\_F2}(P,{Q}^{2}+{U}^{2})
\]
where ${\it \_Fi}$ are  arbitrary functions.

     At the condition
     \[
     b=9\,\sigma-6
\]
this solution takes a more simple form.

In the case
\[
\psi(P,Q,U,V)=H(P,Q)+V
\]
we get the solution
\[
H(P,Q)={\it \_F2}(P)+{\it \_F1}(Q)+{\frac {\left
(1/10\,b-2/5+1/10\, \sigma\right )PQ}{U}}
\]
for which the relation
\[
b=4-\sigma
\]
is a special.

\section{Eikonal equation}

Solution of the eikonal equation
 \begin{equation}\label{dryuma:eq22}
g{{^i}{^j}}\frac{\partial F}{\partial x^i}\frac{\partial
F}{\partial x^j}=0 \end{equation}
 also give us useful information
about the properties of Riemann space.

    In the case of the space with the metric (\ref{dryuma:eq20}) we get the equation
    \[
    2\,\left ({\frac {\partial }{\partial x}}\phi\right )
{\frac {\partial }{\partial P}}\phi+2\,\left ({\frac {\partial
}{\partial y}}\phi\right ){\frac {\partial }{\partial
Q}}\phi+2\,\left ({\frac {\partial }{
\partial z}}\phi\right ){\frac {\partial }{\partial U
}}\phi+2\,\left ({\frac {\partial }{\partial u}}\phi\right ){\frac
{\partial }{\partial V}}\phi-2\,U\left ({\frac {\partial
}{\partial P}}\phi \right ){\frac {\partial }{\partial
Q}}\phi+\]\[+2\,Q \left ({\frac {\partial }{\partial P}}\phi\right
){ \frac {\partial }{\partial U}}\phi+8/5\,\left ({ \frac
{\partial }{\partial P}}\phi\right )\left ({ \frac {\partial
}{\partial V}}\phi\right )P\sigma-2/5 \,\left ({\frac {\partial
}{\partial P}}\phi\right ) \left ({\frac {\partial }{\partial
V}}\phi\right )Pb-2/5\,\left ({\frac {\partial }{\partial P}}\phi
\right )\left ({\frac {\partial }{\partial V}}\phi \right
)P-\]\[-2\,\left ({\frac {\partial }{\partial P}}\phi\right )\left
({\frac {\partial }{\partial V}}\phi \right )rQ-2\,\sigma\,P\left
({\frac {\partial }{\partial Q}}\phi\right ){\frac {\partial
}{\partial V}}\phi+8/5\,\left ({\frac {\partial }{\partial U}}\phi
\right )\left ({\frac {\partial }{\partial V}}\phi \right
)Q-2/5\,\left ({\frac {\partial }{\partial U}}\phi\right )\left
({\frac {\partial }{\partial V}}\phi\right
)Q\sigma-\]\[-2/5\,\left ({\frac {\partial }{\partial
U}}\phi\right )\left ({\frac {\partial }{\partial V}}\phi\right
)Qb+8/5\,\left ({\frac {\partial }{\partial U}}\phi\right )\left
({\frac {\partial }{\partial V}}\phi\right )Ub-2/5\,\left ({\frac
{\partial }{\partial U}}\phi\right )\left ({\frac {\partial
}{\partial V}}\phi\right )U\sigma-2/5\,\left ({\frac {\partial
}{\partial U} }\phi\right )\left ({\frac {\partial }{\partial V}}
\phi\right )U-\]\[-2/5\,V\left ({\frac {\partial }{
\partial V}}\phi\right )^{2}b-2/5\,V\left ({\frac {
\partial }{\partial V}}\phi\right )^{2}\sigma-2/5\,V
\left ({\frac {\partial }{\partial V}}\phi\right )^{2 }=0.
\]

    In particular case
\[
\phi(x,y,z,u,P,Q,U,V)=A(Q,U,V)+PU
\]
one get the equations to determination of the function $A(Q,U,V)$
\[
-2\,{U}^{2}{\frac {\partial }{\partial Q}}A(Q,U,V)+2\,QU{\frac {
\partial }{\partial U}}A(Q,U,V)+2\,QUP+6/5\,U\left ({\frac {\partial }
{\partial V}}A(Q,U,V)\right )P\sigma+6/5\,U\left ({\frac {\partial
}{
\partial V}}A(Q,U,V)\right )Pb-\]\[-4/5\,U\left ({\frac {\partial }{
\partial V}}A(Q,U,V)\right )P-2\,U\left ({\frac {\partial }{\partial V
}}A(Q,U,V)\right )rQ-2\,\sigma\,P\left ({\frac {\partial
}{\partial Q} }A(Q,U,V)\right ){\frac {\partial }{\partial
V}}A(Q,U,V)+\]\[+8/5\,\left ({ \frac {\partial }{\partial
U}}A(Q,U,V)\right )\left ({\frac {\partial }{\partial
V}}A(Q,U,V)\right )Q+8/5\,\left ({\frac {\partial }{
\partial V}}A(Q,U,V)\right )QP-\]\[-2/5\,\left ({\frac {\partial }{
\partial U}}A(Q,U,V)\right )\left ({\frac {\partial }{\partial V}}A(Q,
U,V)\right )Q\sigma-2/5\,\left ({\frac {\partial }{\partial
V}}A(Q,U,V )\right )Q\sigma\,P-\]\[-2/5\,\left ({\frac {\partial
}{\partial U}}A(Q,U,V )\right )\left ({\frac {\partial }{\partial
V}}A(Q,U,V)\right )Qb-2/5 \,\left ({\frac {\partial }{\partial
V}}A(Q,U,V)\right )QbP+\]\[-8/5\, \left ({\frac {\partial
}{\partial U}}A(Q,U,V)\right )\left ({\frac {
\partial }{\partial V}}A(Q,U,V)\right )Ub-2/5\,\left ({\frac {
\partial }{\partial U}}A(Q,U,V)\right )\left ({\frac {\partial }{
\partial V}}A(Q,U,V)\right )U\sigma-\]\[-2/5\,\left ({\frac {\partial }{
\partial U}}A(Q,U,V)\right )\left ({\frac {\partial }{\partial V}}A(Q,
U,V)\right )U-2/5\,V\left ({\frac {\partial }{\partial V}}A(Q,U,V)
\right )^{2}b-2/5\,V\left ({\frac {\partial }{\partial V}}A(Q,U,V)
\right )^{2}\sigma-\]\[-2/5\,V\left ({\frac {\partial }{\partial
V}}A(Q,U,V )\right )^{2} =0
\]
    The solution of this equation is
    \[
A(Q,U,V)=\]\[={\it \_F00}(U){\it \_c}_{{3}}V+1/2\,{\frac
{{Q}^{2}U}{{\it \_c}_{{3}}{ \it
\_F00}(U)\sigma}}+3/5\,QU+3/5\,{\frac {bUQ}{\sigma}}-2/5\,{\frac {
QU}{\sigma}}+2/5\,{\frac {{Q}^{2}}{\sigma}}-1/10\,{Q}^{2}-1/10\,{
\frac {{Q}^{2}b}{\sigma}}+{\it \_F5}(U)
\]
where
\[
{\it \_F00}(U)=5\,{\frac {U}{-4\,{\it \_c}_{{3}}+\sigma\,{\it
\_c}_{{3 }}+{\it \_c}_{{3}}b}},
\]
\[
  {\it \_F5}(U)=1/2\,{\frac {{U}^{2}\left (-2+\sigma\right )}{1+\sigma}}
+{\it \_C1}
\]
 and
\[
b=2/3+2/3\,\sigma,\quad r=1/3+1/3\,\sigma .
\]

 \end{document}